\documentclass[twocolumn,showpacs,floatfix,superscriptaddress]{revtex4}
\usepackage{graphicx}% Include figure files
\usepackage{bm}% bold math

\begin{document}

\title{Thermodynamics of a Trapped Bose-Fermi Mixture}
\author{Hui Hu}
\affiliation{\ Abdus Salam International Center for Theoretical
Physics, P. O. Box 586, Trieste 34100, Italy}
\author{Xia-Ji Liu}
\affiliation{\ LENS, Universit\`{a} di Firenze, Via Nello Carrara
1, 50019 Sesto Fiorentino, Italy} \affiliation{\ Institute of
Theoretical Physics, Academia Sinica, Beijing 100080, China}
\date{\today}

\begin{abstract}
By using the Hartree-Fock-Bogoliubov equations within the Popov approximation,
we investigate the thermodynamic properties of a dilute binary Bose-Fermi
mixture confined in an isotropic harmonic trap. For mixtures with an attractive
Bose-Fermi interaction we find a sizable enhancement of the condensate fraction and
of the critical temperature of Bose-Einstein condensation with respect to the
predictions for a pure interacting Bose gas. Conversely, the influence of the repulsive
Bose-Fermi interaction is less pronounced. The possible relevance of our results
in current experiments on trapped $^{87}{\rm Rb}-^{40}${\rm K} mixtures is discussed.
\end{abstract}

\maketitle

The recent experimental realization of ultracold trapped Bose-Fermi (BF)
mixtures of alkali-metal atoms introduces an interesting new instance of a
quantum many-body system \cite{hulet,ketterle,inguscio}, and has also
stimulated a number of theoretical investigations that address, for example,
the static property \cite{molmer}, the phase diagram and phase separation
\cite{nygaard}, stability conditions \cite{roth} and collective excitations
\cite{japan,capuzzi,liu} of trapped BF mixtures. These investigations have
mainly concentrated at zero temperature using the standard Gross-Pitaevskii
(GP) equation for the Bose gas, in which all the bosonic atoms are assumed
to be in the Bose condensate. An extension of these theories at finite
temperatures where the condensate is strongly depleted is therefore of high
interest, and will also have practical applications. In the theory for a
pure Bose gas, the simplest generalization of the GP equation including the
effect of the noncondensed atoms in a self-consistent manner is the Popov
version of the Hartree-Fock-Bogoliubov (HFB) approximation \cite{griffin}.
As discussed in Ref. \cite{griffin}, this approximation is expected to be
good for both low and high temperatures.

In this paper, we generalize the HFB-Popov approximation to binary BF
mixtures and address the question of how the BF interaction affects the
thermodynamic properties of mixtures. We calculate {\em self-consistently}
the temperature-dependent density profiles of mixtures, as well as the
condensate fraction and the critical temperature of Bose-Einstein
condensation (BEC), at various BF interaction strengths. Our present results
provide the first self-consistent calculation of these thermodynamic
quantities within the HFB-Popov theory which goes beyond the semiclassical
approximation used previously for determining the critical temperature \cite
{albus,ma} and density profiles of binary BF mixtures \cite{tosi}. Our
calculations also show a highly {\em nonlinear} dependence of these
quantities on the BF interaction. In the presence of the BF attraction, the
thermal depletion of the condensate is remarkably decreased and the critical
temperature is shifted towards high temperatures. Conversely, the repulsive
BF interaction affects the condensate fraction and critical temperature in
the opposite direction. However, its influence is less pronounced compared
to the attractive case.

The HFB-Popov mean-field theory for an inhomogeneous interacting Bose gas
has been derived in detail by Griffin in Ref. \cite{griffin}. The
generalization of this theory to a dilute binary BF mixture is
straightforward. Here we merely give a brief summary of basic equations and
emphasize the necessary modification in the presence of the BF interaction.
The trapped dilute mixture is portrayed as a thermodynamic equilibrium
system under the grand canonical ensemble whose thermodynamic variables are $%
N_b$ and $N_f$, respectively, the total number of trapped bosonic and
fermionic atoms, $T$, the absolute temperature, and $\mu _b$ and $\mu _f$,
the chemical potentials. The density Hamiltonian of the system is given by
(in units of $\hbar =1$)
\begin{eqnarray}
{\cal H} &=&{\cal H}_b+{\cal H}_f+{\cal H}_{bf},  \nonumber \\
{\cal H}_b &=&\psi ^{+}({\bf r})\left[ -\frac{{\bf \bigtriangledown }^2}{2m_b%
}+V_{trap}^b({\bf r})-\mu _b\right] \psi ({\bf r})+\frac{g_{bb}}2\psi
^{+}\psi ^{+}\psi \psi ,  \nonumber \\
{\cal H}_f &=&\phi ^{+}({\bf r})\left[ -\frac{{\bf \bigtriangledown }^2}{2m_f%
}+V_{trap}^f({\bf r})-\mu _f\right] \phi ({\bf r}),  \nonumber \\
{\cal H}_{bf} &=&g_{bf}\psi ^{+}({\bf r})\psi ({\bf r})\phi ^{+}({\bf r}%
)\phi ({\bf r}),  \label{hami}
\end{eqnarray}
where $\psi ({\bf r})$ ($\phi ({\bf r})$) is the Bose (Fermi) field operator
that annihilates an atom at position ${\bf r}$. Here we consider a
spherically symmetric system, with trap potentials $V_{trap}^{b,f}({\bf r}%
)=m_{b,f}\omega _{b,f}^2r^2/2$, where $m_{b,f}$ are the atomic masses and $%
\omega _{b,f}$ are the trap frequencies. The interaction between bosons and
between bosons and fermions are described by the contact potentials and are
parameterized by the coupling constants $g_{bb}=4\pi \hbar ^2a_{bb}/m_b$ and $%
g_{bf}=2\pi \hbar ^2a_{bf}/m_r$ in terms of the $s$-wave scattering lengths $%
a_{bb}$ and $a_{bf}$, with $m_r=m_bm_f/(m_{b+}m_f)$ being the
reduced mass. We neglect here the fermion-fermion interactions,
since we are considering a spin-polarized Fermi gas where $s$-wave
collisions are forbidden by the Pauli principle. In the dilute
regime, we may treat the density Hamiltonian describing the BF
coupling in a self-consistent mean-field manner, namely,
\[
{\cal H}_{bf}\simeq g_{bf}\left[ \psi ^{+}\psi \langle \phi ^{+}\phi \rangle
+\langle \psi ^{+}\psi \rangle \phi ^{+}\phi -\langle \psi ^{+}\psi \rangle
\langle \phi ^{+}\phi \rangle \right] .
\]
This kind of decomposition has been used extensively for theoretical
investigations of BF mixtures at zero temperature \cite{note}.

To the Bose field operator $\psi ({\bf r},t)$, we shall apply the usual
decomposition into a $c$-number part plus an operator with vanishing
expectation value: $\psi ({\bf r},t)=\Phi ({\bf r})e^{-i(\varepsilon _0-\mu
_b)t}+\tilde{\psi}({\bf r})$. $\Phi ({\bf r})$ represents the condensate
wave function with eigenvalue $\varepsilon _0$ and the operator $\tilde{\psi}%
({\bf r})$ represents the excitations of the condensate. This ansatz is then
inserted in the equation of motion for $\psi ({\bf r},t)$:
\begin{equation}
i\frac{\partial \psi }{\partial t}=\left[ -\frac{{\bf \bigtriangledown }^2}{%
2m_b}+V_{trap}^b-\mu _b\right] \psi +g_{bb}\psi ^{+}\psi \psi +g_{bf}\langle
\phi ^{+}\phi \rangle \psi .  \label{eom}
\end{equation}
The statistical average over Eq. (\ref{eom}) and the replacement of the
cubic term $\tilde{\psi}^{+}\tilde{\psi}\tilde{\psi}$ by the average in the
mean-field approximation $2\langle \tilde{\psi}^{+}\tilde{\psi}\rangle
\tilde{\psi}$ with neglecting the anomalous expectation value $\langle
\tilde{\psi}\tilde{\psi}\rangle $ and its complex conjugate lead to the
generalized GP equation,
\begin{equation}
{\cal L}_{GP}\Phi ({\bf r})=0,  \label{GPE}
\end{equation}
where ${\cal L}_{GP}\equiv -{\bf \bigtriangledown }^2/2m_b+V_{trap}^b-%
\varepsilon _0+g_{bb}(n_c({\bf r})+2\tilde{n}({\bf r}))+g_{bf}n_f({\bf r})$
with the local density of the condensate $n_c({\bf r})=\left| \Phi ({\bf r}%
)\right| ^2$, of the depletion $\tilde{n}({\bf r})=\langle \tilde{\psi}^{+}(%
{\bf r},t)\tilde{\psi}({\bf r},t)\rangle $, and of the Fermi gas $n_f({\bf r}%
)=\langle \phi ^{+}({\bf r},t)\phi ({\bf r},t)\rangle $. The condensate wave
function in Eq. (\ref{GPE}) is normalized to $N_c=1/(e^{\beta (\varepsilon
_0-\mu _b)}-1)$ with $\beta =(k_BT)^{-1}$.

The subtraction of Eq. (\ref{GPE}) from Eq. (\ref{eom}) gives rise to two
coupled equations of motion for $\tilde{\psi}({\bf r},t)$ and its adjoint,
which can be solved by the usual Bogoliubov transformation, $\tilde{\psi}(%
{\bf r},t)=\sum_i\left[ u_i({\bf r})\hat{\alpha}_ie^{-i\epsilon _it}+v_i^{*}(%
{\bf r})\hat{\alpha}_i^{+}e^{i\epsilon _it}\right] $, to the new Bose
operators $\hat{\alpha}_i$ and $\hat{\alpha}_i^{+}$. This gives the coupled
Bogoliubov-deGennes (BdG) equations,
\begin{eqnarray}
\left[ {\cal L}_{GP}+g_{bb}n_c({\bf r})\right] u_i({\bf r})+g_{bb}n_c({\bf r}%
)v_i({\bf r}) &=&\epsilon _iu_i({\bf r}),  \nonumber \\
\left[ {\cal L}_{GP}+g_{bb}n_c({\bf r})\right] v_i({\bf r})+g_{bb}n_c({\bf r}%
)u_i({\bf r}) &=&-\epsilon _iv_i({\bf r}).  \label{BdG}
\end{eqnarray}
These equations define the quasiparticle excitation energies $\epsilon _i$
{\em relative} to the condensate eigenvalue $\varepsilon _0$, and the
quasiparticle amplitudes $u_i$ and $v_i$. Once these quantities have been
determined, the density of the depletion is obtained in terms of the thermal
number of quasiparticles $\langle \hat{\alpha}_i^{+}\hat{\alpha}_i\rangle
=(ze^{\beta \epsilon _i}-1)^{-1}$ by
\begin{eqnarray}
\tilde{n}({\bf r}) &=&\sum\limits_i\tilde{n}_i({\bf r})\Theta
(E_c^b-\epsilon _i)+\int\limits_{E_c^b}^\infty d\epsilon \tilde{n}(\epsilon ,%
{\bf r}),  \label{dstynt} \\
\tilde{n}_i({\bf r}) &=&\frac{\left| u_i({\bf r})\right| ^2+\left| v_i({\bf r%
})\right| ^2}{ze^{\beta \epsilon _i}-1}+\left| v_i({\bf r})\right| ^2,
\nonumber \\
\tilde{n}(\epsilon ,{\bf r}) &=&\frac{m_b^{3/2}}{\sqrt{2}\pi ^2}\left\{
\frac 1{ze^{\beta \epsilon }-1}+\frac 12-\frac \epsilon {2\epsilon _{HF}}%
\right\}   \nonumber \\
&&\times (\epsilon _{HF}-V_{trap}^b+\varepsilon _0-2g_{bb}n_b({\bf r}%
)-g_{bf}n_f({\bf r}))^{1/2},  \nonumber
\end{eqnarray}
where $z=e^{\beta (\varepsilon _0-\mu _b)}=1+1/N_c$, $\epsilon
_{HF}=(\epsilon ^2+g_{bb}^2n_c^2({\bf r}))^{1/2}$ and $n_b({\bf r})=n_c({\bf %
r})+\tilde{n}({\bf r})$ is the total density of the Bose gas. In the above
equations, to eliminate the numerical errors due to the necessary truncation
of the numerical basis set, we adopt the strategy of Ref. \cite{reidl} and
introduce an energy cutoff $E_c^b$, above which the semiclassical
local-density approximation has been employed.

To solve the generalized GP and BdG equations, one has to find the local
density of the Fermi gas $n_f({\bf r})$. To this end, we insert $\phi ({\bf r%
},t)=\sum_i\varphi _i({\bf r})\hat{c}_ie^{-i\epsilon _it}$ in Eq. (\ref{hami}%
) to diagonalize the quadratic Hamiltonian for $\phi ({\bf r,t})$ in terms
of the new Fermi operator $\hat{c}_i$ that annihilates a fermion at state $%
\varphi _i({\bf r})$. This leads to a stationary Schr\"{o}dinger equation
for $\varphi _i({\bf r})$,
\begin{equation}
\left[ -\frac{{\bf \bigtriangledown }^2}{2m_f}+V_{trap}^f+g_{bf}(n_c+\tilde{n%
})\right] \varphi _i=\epsilon _i\varphi _i.  \label{dfg}
\end{equation}
The density of the Fermi gas is thus obtained by
\begin{eqnarray}
n_f({\bf r}) &=&\sum\limits_in_{fi}({\bf r})\Theta (E_c^f-\epsilon
_i)+\int\limits_{E_c^f}^\infty d\epsilon n_f(\epsilon ,{\bf r}),
\label{dstynf} \\
n_{fi}({\bf r}) &=&\left| \varphi _i({\bf r})\right| ^2\langle \hat{c}_i^{+}%
\hat{c}_i\rangle ,  \nonumber \\
n_f(\epsilon ,{\bf r}) &=&\frac{m_f^{3/2}}{\sqrt{2}\pi ^2}\frac 1{e^{\beta
(\epsilon -\mu _f)}+1}(\epsilon -V_{trap}^f-g_{bf}n_b({\bf r}))^{1/2},
\nonumber
\end{eqnarray}
where $\langle \hat{c}_i^{+}\hat{c}_i\rangle =(e^{\beta (\epsilon _i-\mu
_f)}+1)^{-1}$ is the Fermi distribution. Analogously to Eq. (\ref{dstynt})
we have applied the finite-temperature Thomas-Fermi (TF) approximation only
for high-lying Fermi levels above an energy cutoff $E_c^f$ to avoid the
truncation errors.

Equations (\ref{GPE})-(\ref{dstynf}) form a closed system of equations that
we have referred to as the ``HFB-Popov'' equations for a dilute BF mixture.
We have numerically solved these equations by an iterative procedure as
follows: The generalized GP and BdG equations are first solved
self-consistently for $\Phi ({\bf r})$, $u_i({\bf r})$, and $v_i({\bf r})$
as described in Ref. \cite{griffin} to evaluate $n_c({\bf r})$ and $\tilde{n}%
({\bf r})$, with $n_f({\bf r})$ set to the result for an ideal Fermi gas.
Once $n_c({\bf r})$ and $\tilde{n}({\bf r})$ are known, the eigenfunctions
in Eq. (\ref{dfg}) are obtained numerically and are used to update $n_f({\bf %
r})$ in Eq. (\ref{dstynf}). This newly generated $n_f({\bf r})$ is then
inserted in the GP and BdG equations and the process is iterated to
convergence. At each step, the chemical potentials for the Bose gas and the
Fermi gas are fixed by the normalization conditions, $\int d{\bf r}n_b({\bf r%
})=N_b$ and $\int d{\bf r}n_f({\bf r})=N_f$, respectively.

As an illustration of this procedure, we consider a mixture of 2000 $^{87}%
{\rm Rb}$ (boson) and 2000 $^{40}${\rm K} (fermion) atoms in an isotropic
harmonic trap, for which the order parameter $\Phi ({\bf r})$, the
quasiparticle amplitudes $u_i({\bf r})$ and $v_i({\bf r})$, and the orbits $%
\varphi _i({\bf r})$ can be labelled by $(n,l,m)$, according to
the number of nodes in the radial solution $n$, the orbital
angular momentum $l$, and its
projection $m$. In addition, we use the following parameters \cite{inguscio}%
: $m_b=1.45\times 10^{-25}$ {\rm kg}, $\omega _b=2\pi \times 216$ {\rm Hz}, $%
m_f/m_b=0.463$, $\omega _f/\omega _b=1.47$, $a_{bb}=99a_0$, and $%
a_{bf}=-410a_0$, where $a_0=0.529$ ${\rm \AA }$ is the Bohr radius. Because
our calculations are especially delicate near the critical temperature, we
have taken $n_{\max }=32$, $l_{\max }=64$ and high energy cutoffs of $%
E_c^b=60\hbar \omega _b$ and $E_c^f=90\hbar \omega _b$ to ensure the
accuracy. Throughout the paper, we also express the lengths and energies in
terms of the characteristic oscillator length $a_{ho}^b=(\hbar /m_b\omega
_b)^{1/2}$ and characteristic trap energy $\hbar \omega _b$, respectively.

\begin{figure}[tbp]
\centerline{\includegraphics[width=7.0cm,angle=-90,clip=]{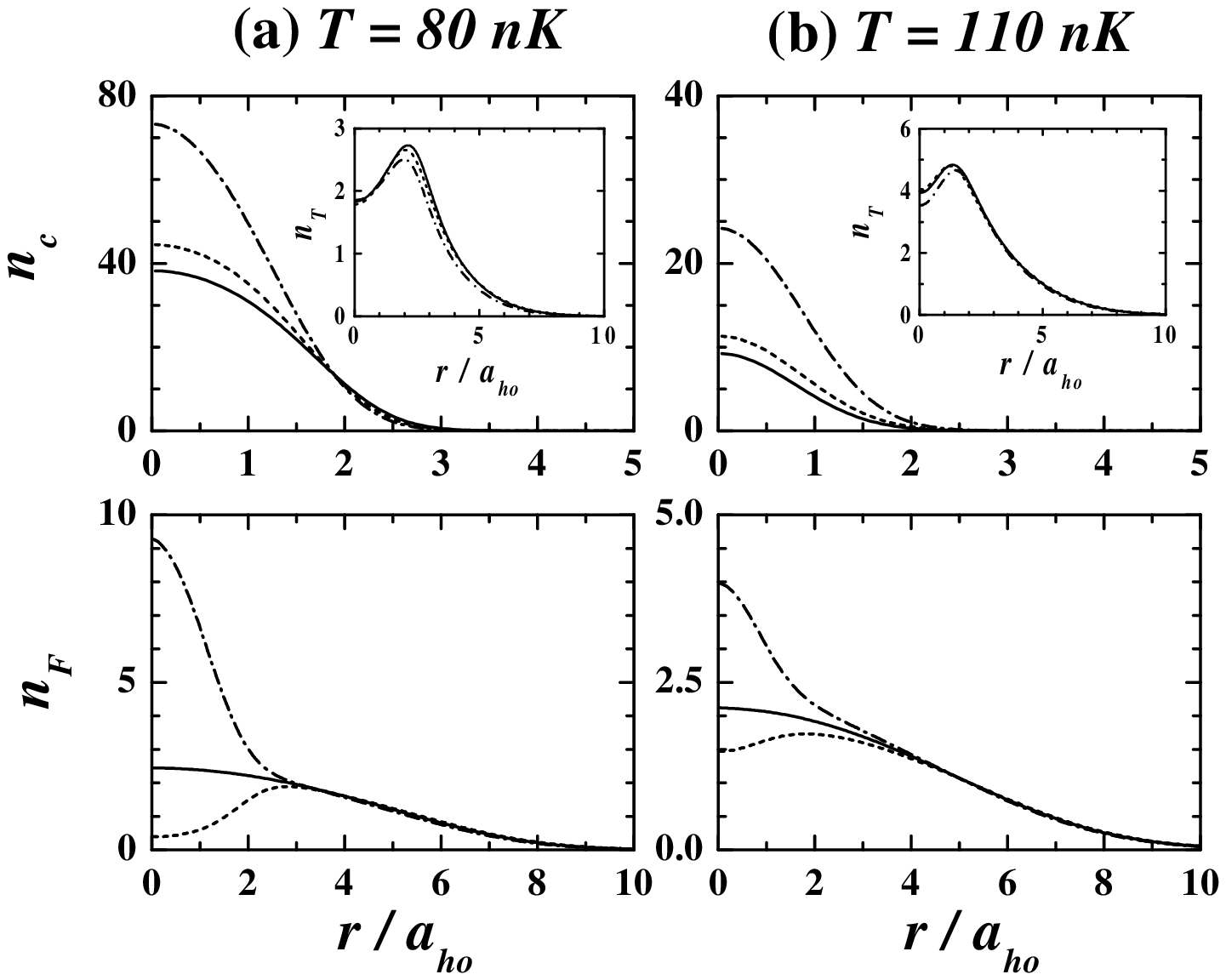}}
\caption{The density profiles of the condensate $n_c(r)$ and of
the Fermi gas $n_F(r)$ for a mixture of 2000 $^{87}{\rm Rb}$ and
2000 $^{40}${\rm K} atoms in an isotropic harmonic trap at various
BF interaction: $a_{bf}=0$ (full line), $a_{bf}=+410a_0$ (dashed
line), and $a_{bf}=-410a_0$ (dash-dotted line), where $a_0=0.529$
${\rm \AA }$ is the Bohr radius. Insets show the density profiles
of the noncondensate $\tilde{n}(r)$. We have taken $a_{bb}=99a_0$
and $\omega _b=2\pi\times 216$ {\rm Hz}. The coordinate $r$ and
densities are measured in units of the harmonic oscillator length
$a_{ho}^b$ and $\left( a_{ho}^b\right) ^3$, respectively.}
\label{fig1}
\end{figure}

In Fig. 1, we present our results for the density profiles of the
condensate, of the noncondensate, and of the Fermi gas at two temperatures.
The cases with the BF interaction and without the BF interaction are shown
by the dash-dotted lines and full lines, respectively. We have also
considered a fictitious case of a positive BF interaction: $a_{bf}=+410a_0$
(dashed lines). The choice of the first temperature, $T=80$ {\rm nK},
corresponds to the situation in which the condensate and noncondensate have
an approximately equal number of atoms, while the other temperature $T=110$
{\rm nK} is chosen to be close to the critical temperature for a pure
interacting Bose gas with the same number of bosons, $T_c\approx 112$ {\rm nK%
}. As clearly emerges from the figure, the density profiles of the
condensate and of the Fermi gas are strongly affected by the BF interaction
at both temperatures. In particular, the densities around the center are
significantly enhanced in the case of the BF attraction. The density profile
of the noncondensate (shown in the insets), on the other hand, is less
influenced by the BF interaction due to its broad distribution and the
strong repulsion from the condensate.

\begin{figure}[tbp]
\centerline{\includegraphics[width=5.5cm,angle=-90,clip=]{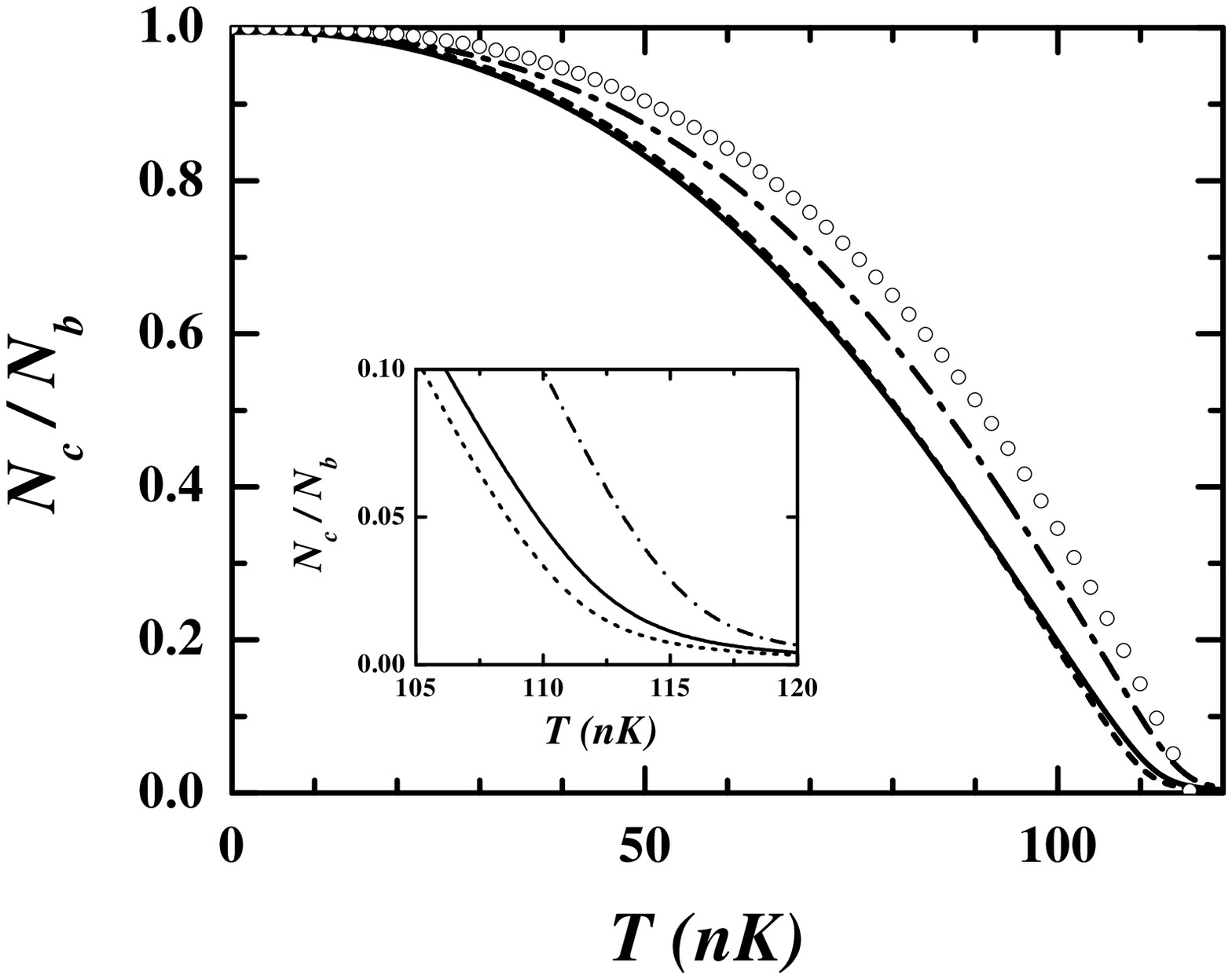}}
\caption{Temperature dependence of the condensate fraction. The
dashed and dash-dotted lines correspond to $a_{bf}=+410a_0$ and
$-410a_0$, respectively. The solid lines are the result for a pure
interacting Bose gas. The empty circles are ideal Bose gas result
with the finite size
correction, $N_c^0/N_b=1-(T/T_c^0)^3-2.1825(T/T_c^0)^2N_b^{-1/3}$, where $%
k_BT_c^0=0.94\hbar \omega _bN_b^{1/3}$. The inset highlights the
condensate fraction near the critical temperature. The other
parameters are the same as in Fig.1.}
\label{fig2}
\end{figure}

In Fig. 2, we show our predictions for the temperature dependence of the
condensate fraction $N_c(T)/N_b$. The essential feature of the figure is the
importance of the {\em attractive} BF interaction that results in a sizable
quenching of the thermal depletion compared to the prediction for a pure
interacting Bose gas. Contrarily, the effects of the repulsive BF
interaction are more subtle and are always very small. The sizable
enhancement of the condensate fraction predicted by our calculation follows
from the fact that in the presence of the BF attraction the condensate
effectively experiences a more tightly confining potential. As a
consequence, if we neglect the corrections due to the interaction between
bosons and the finite size effect, the critical temperature $T_c^0$ $%
=0.94\hbar \omega _{eff}N_b^{1/3}/k_B$ is effectively increased and the
condensate fraction is, therefore, enhanced according to the ideal gas
result $N_c/N_b=1-(T/T_c^0)^3$.

\begin{figure}[tbp]
\centerline{\includegraphics[width=5.5cm,angle=-90,clip=]{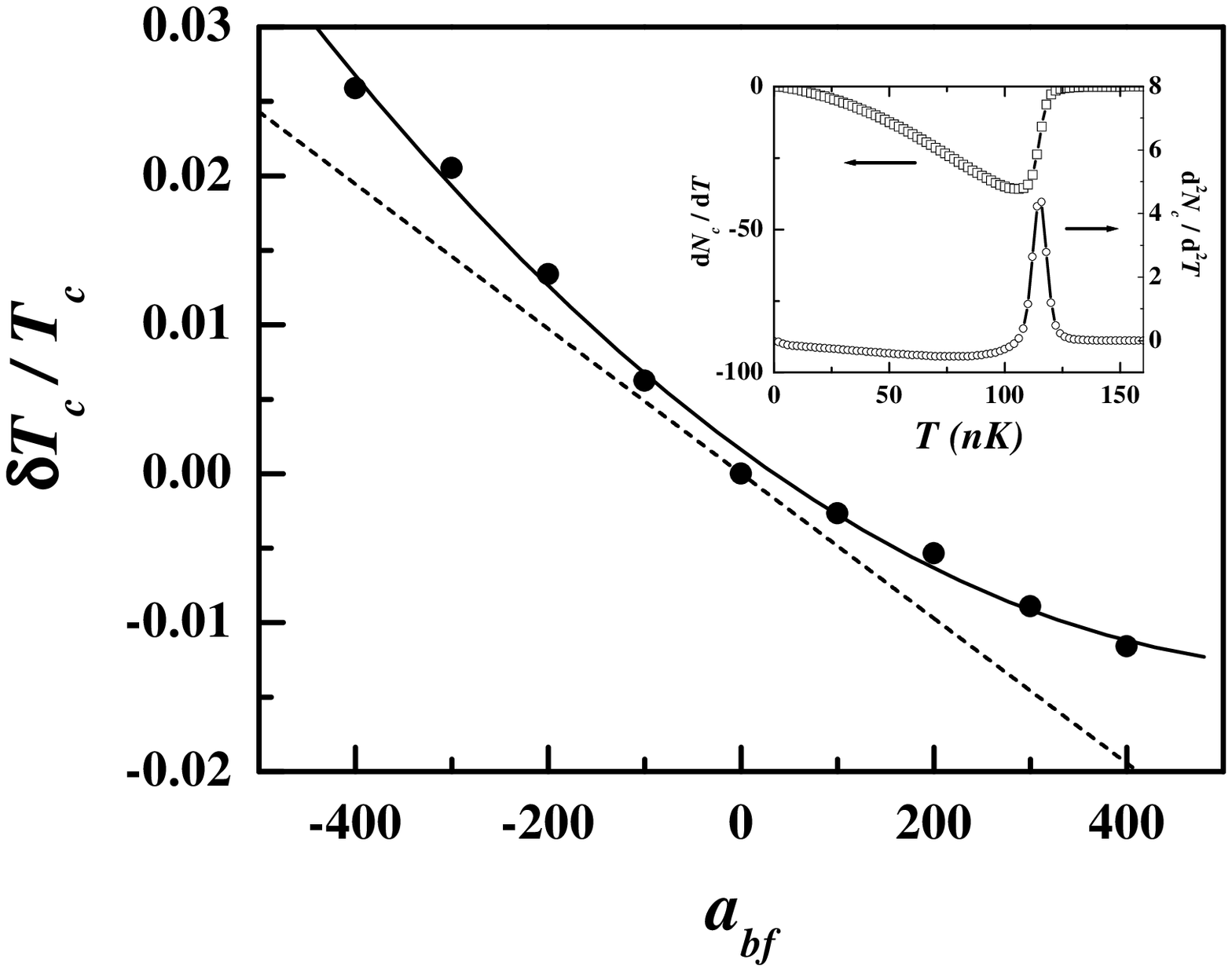}}
\caption{The relative shift of the critical temperature as a
function of the BF $s$-wave scattering length $a_{bf}$ (in units
of $a_0$). The solid circles are the result calculated using the
HFB-Popov equations and the full line is a parabolic fit to guide
eye. The dashed line is the result calculated by Eq. (18) in Ref.
[11]. The inset shows the functions $dN_c/dT$ and $d^2N_c/dT^2$
for the case of $a_{bf}=-410a_0$. $T_c$ is extracted from the
maximum of $d^2N_c/dT^2 $. The other parameters are the same as in
Fig.1. } \label{fig3}
\end{figure}

Closely related to the condensate fraction, another important parameter
characterizing the effect of the BF interaction is the shift of the critical
temperature from the pure interacting Bose gas case. In Fig. 3, we report
the HFB-Popov results for the relative shift of the critical temperature $%
\delta T_c/T_c$ as a function of $a_{bf}$ in solid circles. Here $T_c$ is
determined as the maximum of the function $d^2N_c/dT^2$ \cite{bergeman}. The
semiclassical predictions for $\delta T_c/T_c$, calculated as in Ref. \cite
{albus} in the first order of $a_{bf}$, are also shown by the dashed line.
The agreement of these two approaches is reasonably good for a weak BF
interaction ($\left| a_{bf}\right| $ $\lesssim 100a_0$). However, as $\left|
a_{bf}\right| $ increases, our HFB-Popov results diverge from the
semiclassical predictions. In particular, for the realistic BF $s$-wave
scattering length for $^{87}{\rm Rb}-^{40}${\rm K} mixtures, $a_{bf}=-410a_0$%
, the deviation becomes remarkable.

We now turn to consider the experimental relevance of our results. In
current experiments, the realistic number of $^{87}{\rm Rb}$ and $^{40}${\rm %
K} atoms in mixtures is about ten times larger than what we assumed here
\cite{inguscio}. For such large number of atoms, our calculation is very
time consuming and we then resort to the semiclassical version of the
HFB-Popov theory by setting $E_c^b$ to an energy of a few $\hbar \omega _b$
and applying the TF approximation for the whole Fermi spectra. The accuracy
of this semiclassical treatment has been checked by the comparison with the
full quantum mechanical calculations for a small mixture. The condensate
fraction obtained by these two methods coincides within $1\%$ errors.

\begin{figure}[tbp]
\centerline{\includegraphics[width=5.5cm,angle=-90,clip=]{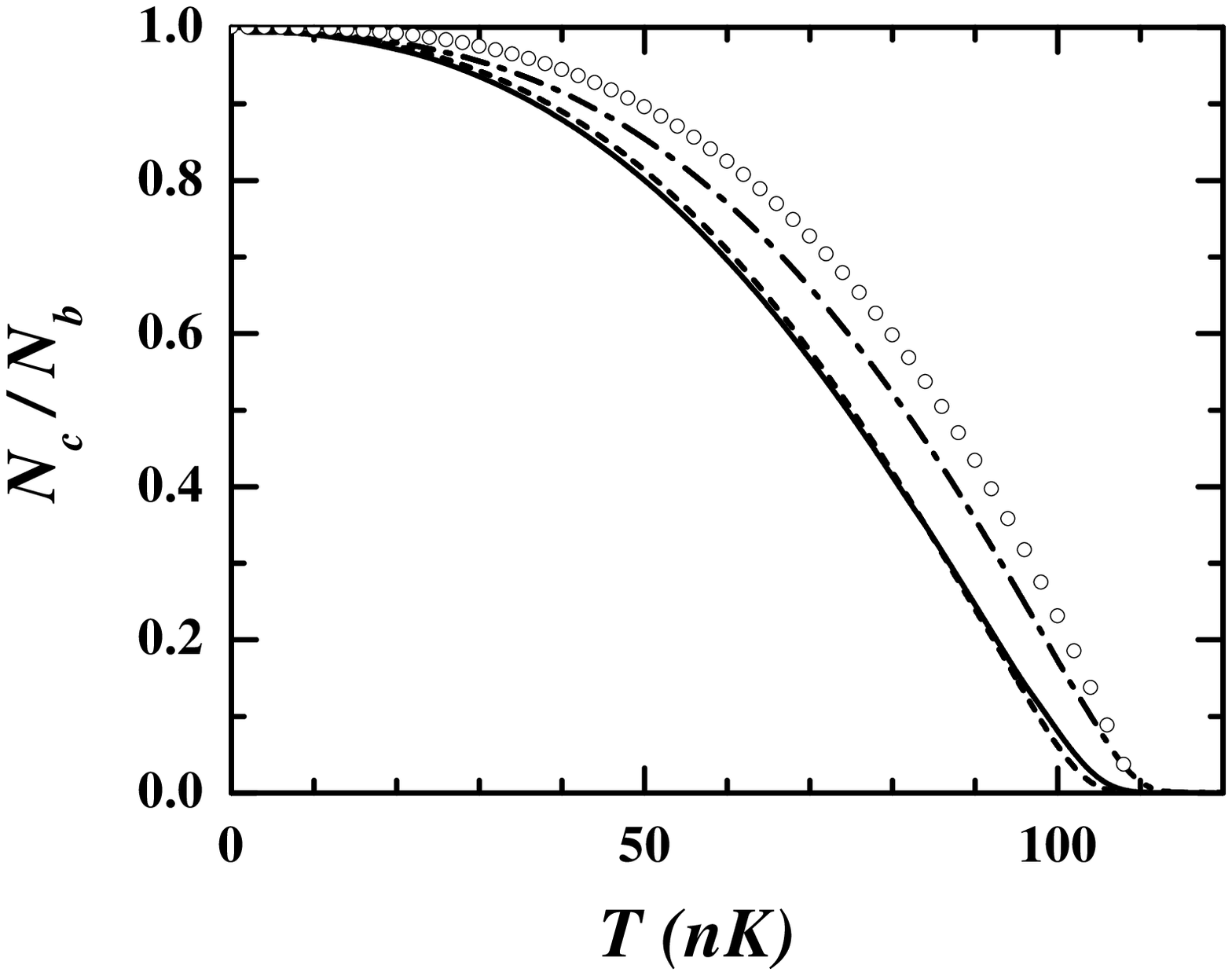}}
\caption{Temperature dependence of the condensate fraction for a mixture of $%
2\times 10^4$ $^{87}{\rm Rb}$ and $5\times 10^4$ $^{40}${\rm K}
atoms confined in an isotropic trap with $\omega _b=2\pi\times
91.7$ {\rm Hz}, calculated in the semiclassical version of the
HFB-Popov theory as mentioned in the text.} \label{fig4}
\end{figure}

In Fig. 4, we present the results for $N_c(T)/N_b$ of a mixture of
$2\times 10^4$ $^{87}{\rm Rb}$ and $5\times 10^4$ $^{40}${\rm K}
atoms confined in an isotropic trap with $\omega _b=2\pi \times
91.7$ {\rm Hz}. This choice corresponds to the typical
experimental situations at LENS \cite{note2}. Although the trap
becomes more shallow, the enhancement of the condensate fraction
shown in Fig. 4 is quantitatively similar to the one of Fig. 2,
due to the much larger values of $N_b$ and $N_f$ contained in the
trap. Finally, in this case we also roughly estimate the relative
shift of the critical temperature due to the BF attraction,
$(\delta T_c/T_c)_{bf}\simeq +4\%$,
which is comparable to the shift due to the boson-boson interaction, $%
(\delta T_c/T_c)_{bb}\simeq -1.33(a_{bb}/a_{ho}^b)N_b^{1/6}=-3.2\%$, and the
finite size correction, $(\delta T_c/T_c)_{fs}\simeq -0.73N_b^{-1/3}=-2.7\%$
\cite{giorgini}.

In conclusion, we have generalized the HFB-Popov theory to binary
BF mixtures and have presented a detailed study of the
thermodynamic properties of mixtures at finite temperature,
including the density profiles, the condensate fraction, as well
as the critical temperature of BEC. These quantities are found to
depend on the BF interaction in a nonlinear way. Moreover, under
conditions appropriate to the $^{87}{\rm Rb}-^{40}${\rm K} mixture
in the LENS experiments, the condensate fraction and the critical
temperature of BEC are significantly enhanced with respect to the
prediction for a pure interacting Bose gas. This enhancement might
be observable in current experiments.\newline

%===============================================================
\begin{acknowledgments}
We are very grateful to Dr. M. Modugno and Dr. G. Modugno for
simulating discussions and for a careful reading of the
manuscript. One of us (H.H.) would like to thank the hospitality
of LENS while part of this work was performed. X.-J.L was
supported by the K.C.Wong Education Foundation, the Chinese
Research Fund, the NSF-China under Grant No. 10205022, and the
''ICTP Programme for Training and Research in Italian Laboratories
(TRIL), Trieste, Italy''.
\end{acknowledgments}
%===============================================================


\begin{thebibliography}{99}

\bibitem{hulet}  A. G. Truscott, K. E. Strecker, W. I. McAlexander, G. B.
Partridge, and R. G. Hulet, Science {\bf 291}, 2570 (2001); F. Schreck, L.
Khaykovich, K. L. Corwin, G. Ferrari, T. Bourdel, J. Cubizolles, and C.
Salomon, Phys. Rev. Lett. {\bf 87}, 080403 (2001).

\bibitem{ketterle}  Z. Hadzibabic, C. A. Stan, K. Dieckmann, S. Gupta, M. W.
Zwierlein, A. Gorlitz, and W. Ketterle, Phys. Rev. Lett. {\bf 88}, 160401
(2002).

\bibitem{inguscio}  G. Roati, F. Riboli, G. Modugno, and M. Inguscio, Phys.
Rev. Lett. {\bf 89}, 150403 (2002); G. Modugno, G. Roati, F. Riboli, F.
Ferlaino, R. J. Brecha, and M. Inguscio, Science {\bf 297}, 2240 (2002).

\bibitem{molmer}  K. M\o lmer, Phys. Rev. Lett. {\bf 80}, 1804 (1998).

\bibitem{nygaard}  N. Nygaard and K. M\o lmer, Phys. Rev. A {\bf 59}, 2974
(1999).

\bibitem{roth}  R. Roth and H. Feldmeier, Phys. Rev. A {\bf 65}, 021603(R)
(2002).

\bibitem{japan}  T. Miyakawa, T. Suzuki, and H. Yabu, Phys. Rev. A {\bf 62},
063613 (2000).

\bibitem{capuzzi}  P. Capuzzi and E. S. Hern\'{a}ndez, Phys. Rev. A {\bf 64}%
, 043607 (2001).

\bibitem{liu}  X.-J. Liu and H. Hu, Phys. Rev. A {\bf 67}, 023613 (2003).

\bibitem{griffin}  A. Griffin, Phys. Rev. B 53, 9341 (1996); D. A.
Hutchinson, E. Zaremba, and A. Griffin, Phys. Rev. Lett. {\bf 78}, 1842
(1997).

\bibitem{albus}  A. P. Albus, S. Giorgini, F. Illuminati, and L. Viverit, J.
Phys. B {\bf 35}, L511 (2002).

\bibitem{ma}  Y.-L. Ma and S.-T. Chui, Phys. Rev. A {\bf 66}, 053611 (2002).

\bibitem{tosi}  M. Amoruso, A. Minguzzi, S. Stringari, M. P. Tosi, and L.
Vichi, Eur. Phys. J. D {\bf 4}, 261 (1998).

\bibitem{albus2}  A. P. Albus, S. A. Gardiner, F. Illuminati, and M.
Wilkens, Phys. Rev. A {\bf 65}, 053607 (2002).

\bibitem{viverit}  L. Viverit and S. Giorgini, Phys. Rev. A {\bf 66}, 063604
(2002).

\bibitem{albus3}  A. P. Albus, F. Illuminati, and M. Wilkens, Phys. Rev. A
{\bf 67}, 063606 (2003).

\bibitem{note}  The high-order correlation beyond this mean-field
decomposition has been considered at zero temperature for a homogeneous BF
mixture in Refs. \cite{albus2,viverit}, and for a trapped BF mixture in Ref.
\cite{albus3} by applying the local-density approximation.

\bibitem{reidl}  J. Reidl, A. Csord\'{a}s, R. Graham, and P. Sz\'{e}pfalusy,
Phys. Rev. A {\bf 59}, 3816 (1999).

\bibitem{bergeman}  T. Bergeman, D. L. Feder, N. L. Balazs, and B. I.
Schneider, Phys. Rev. A {\bf 61}, 063605 (2000).

\bibitem{note2}  The trap frequency taken here is the geometrical average of
the frequencies of the axially symmetric trap at LENS. As checked in Ref.
\cite{giorgini}, the effects of the deformation of the trap on the
condensate fraction are always very small for the large values of $N_b$.

\bibitem{giorgini}  S. Giorgini, L. P. Pitaevskii, and S. Stringari, Phys.
Rev. A {\bf 54}, R4633 (1996).

\end{thebibliography}
\end{document}